\documentclass[10pt]{article}
\usepackage[OE]{express}

\usepackage{graphicx,subfigure,color,amssymb,amsmath,mathtools,bm}

\newcommand{\openone}{\leavevmode\hbox{\small1\normalsize\kern-.rea33em1}}
\newcommand{\sugg}[1]{{#1}}
\newcommand{\re}{\operatorname{Re}}
\newcommand{\Vr}{\mathbf{r}}
\newcommand{\vrho}{\bm{\rho}}
\newcommand{\vk}{\mathbf{k}}
\newcommand{\vka}{\bm{\kappa}}

\newcommand{\dd}{{d^{\,2}}}

\providecommand{\norm}[1]{\lVert#1\rVert}

\begin{document}
\title{Unraveling beam self-healing}

\author{Andrea~Aiello,{\authormark{1,2}} 
 Girish~S.~Agarwal,{\authormark{3,4}}
 Martin~Pa\'{u}r,{\authormark{5}} 
Bohumil~Stoklasa,{\authormark{5}}
Zden\v{e}k~Hradil,{\authormark{5}}
Jaroslav~\v{R}eh\'{a}\v{c}ek,{\authormark{5}} 
Pablo~de~la~Hoz,{\authormark{6}} 
Gerd~Leuchs,{\authormark{1}} and 
Luis~L.~S\'anchez-Soto{\authormark{1,6,*}}} 

\address{
\authormark{1}Max Planck Institut f\"ur die Physik des Lichts,
  Staudtstra\ss e 2, 91058 Erlangen, Germany \\
\authormark{2}Institut f\"ur Theoretische Physik II, Friedrich-Alexander 
  Universit\"{a}t Erlangen-N\"{u}rnberg, Staudtstra\ss e 2,
  91058~Erlangen, Germany \\
\authormark{3}Institute for Quantum Science and Engineering 
 and Department of Biological and Agricultural Engineering, 
 Texas A\&M University, College~Station, Texas~77845, USA \\
\authormark{4}Department of Physics, Oklahoma State University,
  Stillwater, Oklahoma 74078, USA \\
\authormark{5}Department of Optics,  Palack\'y  University, 
17.~listopadu 12, 771 46 Olomouc,  Czech Republic\\
\authormark{6}Departamento de \'Optica, Facultad de F\'{\i}sica,
 Universidad Complutense, 28040~Madrid,  Spain} 
\email{\authormark{*}lsanchez@fis.ucm.es} %% email address is required



\begin{abstract}
  We show that, contrary to popular belief, non only diffraction-free
  beams may reconstruct themselves after hitting an opaque obstacle
  but also, for example, Gaussian beams.  We unravel the mathematics
  and the physics underlying the self-reconstruction mechanism and we
  provide for a novel definition for the minimum reconstruction
  distance beyond geometric optics, which is in principle applicable
  to any optical beam that admits an angular spectrum representation.
  Moreover, we propose to quantify the self-reconstruction ability of
  a beam via a newly established degree of self-healing. This is
  defined via a comparison between the amplitudes, as opposite to
  intensities, of the original beam and the obstructed one. Such
  comparison is experimentally accomplished by tailoring an innovative
  experimental technique based upon Shack-Hartmann wave front
  reconstruction.  We believe that these results can open new avenues
  in this field.
\end{abstract}


%%%%%%%%%%%%%%%%%%%%%%% References %%%%%%%%%%%%%%%%

%%%%%%%%%%%%%%%%%%%%%%%%%%  body  %%%%%%%%%%%%%%%%%%%%%%%%%%

\section{Introduction}

In recent years, the remarkable capacity of a beam to reconstruct
itself after encountering an obstacle (frequently called self-healing)
has attracted a good deal of attention~\cite{Bouchal:1998aa,
  Vasnetsov:2000aa,Garces:2004aa} and has already found applications
in diverse areas~\cite{Arlt:2001aa,Garces:2002aa,Fahrbach:2010aa,
  Fahrbach:2012aa,McLaren:2014aa}.

Self-healing has been long time considered as a distinctive feature of
nondiffracting beams~\cite{Durmin:1987aa}; most prominently of Bessel
beams~\cite{Bouchal:2002aa,Tao:2004aa,Fischer:2006aa,Chu:2012aa},
although also Airy~\cite{Broky:2008aa},
caustic~\cite{Anguiano:2007aa}, and Mathieu and
Weber~\cite{Zhang:2012aa} beams have been examined.

It was subsequently realized that some diffracting beams, including
the whole family of scaled propagation invariant
beams~\cite{Arrizon:2015aa} optical ring
lattices~\cite{Vainty:2011aa}, Pearcey beams~\cite{Ring:2012aa}, and
tightly focused~\cite{Vyas:2011aa} and radially
polarized~\cite{Wu:2014aa} Bessel-Gauss beams, can
self-reconstruct. However, there is still the widespread perception
that the self-reconstruction hinges on engineering special beam
profiles and, in many instances, it is sensitive to the obstruction
size and shape, thereby limiting applications of this
phenomenon~\cite{Wang:2016aa}.

Recently, a complete account of self-healing for Bessel beams has been
given in terms of wave optics~\cite{Aiello:2014aa}. The basic
mechanism can be entirely explained in terms of the propagation of
plane waves with radial wave vectors lying on a ring. The results
obtained are in agreement with the standard ones established from a
geometrical approach~\cite{McGloin:2005aa,Litvin:2009aa}, yet they
open a new scope.

In this paper, still using a wave-optics methodology, we come to the
conclusion that self-healingmay occur, potentially, for almost any
kind of beam. Note, though, that it is outside the scope of most
self-healing researches, and the present work is not an exception, the
study of self-reconstruction capabilities of structured optical beams,
as multiple-beam assemblies and, more generally, beams with complex
and intricate intensity, polarization, frequency and temporal
structures~\cite{Andrews:2008aa}. 

Furthermore, we introduce an appropriate degree that quantifies
the similarity between the field of the unperturbed beam (namely, the
beam that would propagate as if the obstacle were not present) and the
field of the perturbed one (that is, the beam that propagates behind
the obstruction). In this way, we put in evidence that self-healing is a
property of both the intensity and the phase of the spatial
distribution of the beam. We experimentally test these issues with
a Gaussian beam, whose intensity and phase are measured by means
of a CCD camera and a Shack-Hartman wavefront sensor, finding an
outstanding agreement with our theoretical predictions.

\section{Self-healing as an eigenvalue problem}
\label{sec:eig}

Let us first set the stage for our construction. We consider a scalar
field $\Psi (x,y,z)$ propagating along the $z$-axis. An obstruction,
characterized by an amplitude transmission function $t_O (x,y)$,
is placed in the plane $z=0$. Here and hereafter with obstruction we
denote any physical object that decreases the intensity of a light
beam, possibly in a space-dependent manner, without changing directly
phase and polarization of light. The amplitude $\Psi_O(x,y,0)$ of the
obstructed field at the plane $z=0$ is
\begin{equation}
  \label{p10}
  \Psi_{O} (x,y,0) =  t_{O}(x,y) \, \Psi (x,y,0) \, .
\end{equation}
The angular spectrum representation~\cite{Mandel:1995aa} is probably
the most germane method to deal with the field propagation.  Accordingly, 
the amplitude $\Psi_{O}(x,y,z)$ of the field transmitted at a distance
$z$ from the obstruction can be expressed as the plane-wave superposition 
\begin{align}
  \label{p35}
  \Psi_{O}(x,y,z) &  =   
  \frac{1}{(2 \pi)^2}
  \iint\limits_{-\infty}^{\phantom{xx}\infty} 
 \exp (i \vrho \cdot \vka )  \exp (i z k_z )   
%\nonumber \\
%& \times 
 \left[ \iint\limits_{-\infty}^{\phantom{xx}\infty} 
\widehat{t}_O(\vka - \vka{^{\prime}}) \widehat{\Psi}(\vka{^{\prime}}) \, 
 \dd  \vka{\,'} \right]  \dd \vka \, .
\end{align}
The wide hat (not to be confused with the small hat marking unit
vectors) will denote throughout the spatial Fourier transform of the
corresponding function evaluated at $z=0$; i.e., its angular
spectrum. Two-dimensional transverse vectors, in either real and
Fourier space, are denoted with Greek letters:
$\vrho = x \hat{x} + y \hat{y}$ and
$\vka = k_x \hat{x} + k_y \hat{y}$. In addition,
$k_z = (k^2 - \kappa^2 )^{1/2}$, with
$\kappa^2 = k_x^2 + k_y^2$.

Given the function $t_O(x,y)$, one can always define the transmission
function $t_A(x,y)$ of an aperture complementary to the
obstruction~\cite{Born:1999aa} via the Babinet principle
$t_A(x,y) + t_O(x,y) =1$. Therefore, \eqref{p10} yields 
\begin{equation}
  \label{p10b}
  \Psi_O(x,y,0)  =   [ 1 -t_A(x,y) ] \Psi (x,y,0) 
  \equiv   \Psi (x,y,0)  - \Psi_A(x,y,0).
\end{equation}
Taking the absolute value squared of both sides of this equation and
integrating over the whole $xy$-plane, we obtain
\begin{equation}
  \label{Lambda}
  \mathcal{I} [ \Psi_O ] = 
  \mathcal{I}[ \Psi ] + \mathcal{I}  [\Psi_A] 
  - 2 \re  \iint\limits_{-\infty}^{\phantom{xx}\infty} 
  \Psi^{\ast} (x,y,0) \Psi_A(x,y,0)  \, dx dy,
\end{equation}
where
$\mathcal{I} [h] = \iint\limits_{-\infty}^{\phantom{xx}\infty}
h^{\ast}(x,y,z) h(x,y,z) \, dx dy$ is the average beam intensity at
the plane $z$.

Conventionally, a beam is dubbed self-healing when it has the ability
to recover its amplitude or intensity profile after being obscured by
an obstacle. Quite obviously, perfect self-reconstruction is
impossible, even in principle, because, as Eq.~(\ref{Lambda})
distinctly shows, the intensity of the transmitted field is
unavoidably reduced unless $\mathcal{I}[\Psi_{A}]=0$.  We thus content
ourselves with the condition
\begin{equation}
\label{p40}
  \Psi_O(x,y,z) \approx   \lambda_{0}  \Psi (x,y, z ) \, , 
  \qquad
  \forall z \geq z_{0} \, ,
\end{equation}
where $z_0$ denotes the so-called minimum reconstruction
  distance  and the scaling factor
$ \lambda_0= \left \{ \mathcal{I}[\Psi_O]/ \mathcal{I}[\Psi] \right
\}^{1/2} $ accounts for the average intensity reduction caused by the 
obstruction. 

The left-hand side of \eqref{p40} is given by (\ref{p35}), while the
field in the right-hand side can be jotted down as
\begin{equation}
  \label{p50}
  \Psi (x, y, z)  =  \frac{1}{2 \pi} \iint\limits_{-\infty}^{\phantom{xx}\infty}
  \exp ( i \vrho \cdot \vka )  
  \exp ( i z k_z ) 
  \widehat{\Psi} (\vka)   \, \dd \kappa. 
\end{equation}
Consequently, \eqref{p40} can be equivalently recast as
\begin{equation}
  \label{p150}
  \widehat{\Psi}_O (\vka)  \approx   \lambda_0 
  \widehat{\Psi} (\vka)   \, .
\end{equation}
Notice carefully, though, that \eqref{p150} does not contain the
variable $z$, whereas the relation (\ref{p40}) is supposed to be valid
only for $z \geq z_0$. The latter requirement cannot be ignored
because \eqref{p40} cannot be satisfied at $z=0$, where instead
\eqref{p10b} must be fulfilled. Of course, in \eqref{p10b} we are
implicitly excluding the trivial case of a spatially-uniform
semi-transparent intensity obstruction (think of, e.g., a
neutral-density filter) such that
$\Psi_A(x,y,0) = (1-\lambda_0) \Psi(x,y,0)$.

Hence, we are apparently faced with a contradiction here. In fact,
\eqref{p150} constitutes more a statement about the obstruction rather
than the field. This can be seen by rewriting \eqref{p150} in the more
enlightening form
\begin{align}
  \label{p180}
  \frac{1}{2 \pi} \iint\limits_{-\infty}^{\phantom{xx}\infty} 
  \widehat{t}_O(\vka - \vka^{\prime}) 
  \widehat{\Psi} (\vka^{\prime}) \, \dd \kappa' 
  \approx \lambda_0   \widehat{\Psi} (\vka)  \, . 
\end{align}
With the equality sign, this is a homogeneous Fredholm integral
equation~\cite{Kress:1999aa} for the function $\widehat{\Psi} (\vka)$,
which has to be an eigenfunction with eigenvalue $2 \pi \lambda_0$, of
the integral kernel $ \widehat{t}_{O} (\vka - \vka^{\prime})$
describing the obstruction.  This means that the requirement
\eqref{p40} is indeed too much restrictive because it can be satisfied
only by those beams whose angular spectrum (the eigenfunction) is
unaffected by the interaction with the obstruction, apart from a
trivial proportionality factor (the eigenvalue), as shown in
\eqref{p180}.

%%%%%%%%%%%%%%%%%%%%%%%%%%%%%%%%%%%%%%%%%%
\begin{figure}[t]
  \centerline{\includegraphics[height=5.5cm]{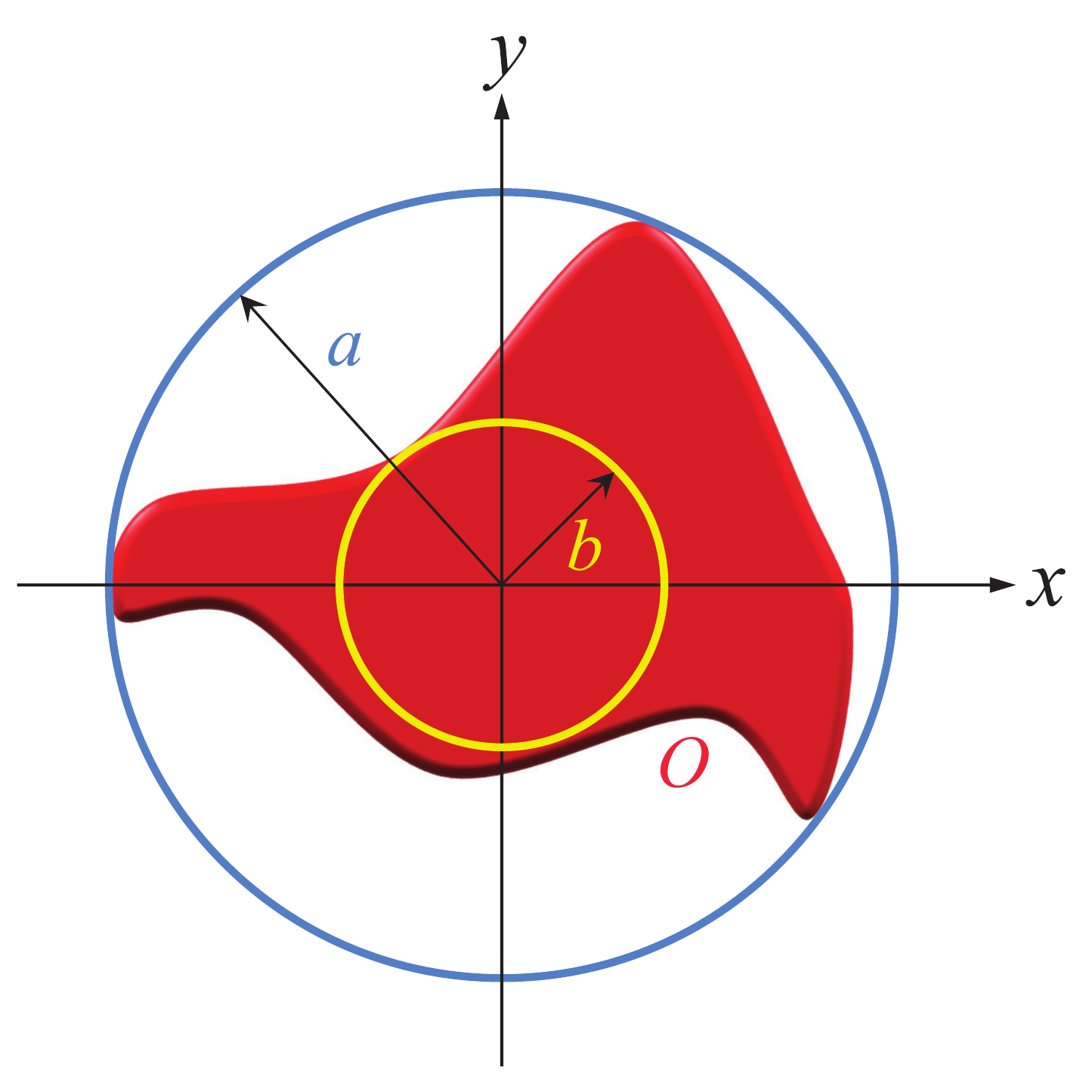}}
  \caption{Obstruction of area $O$ represented in red.  This
    region is circumscribed by the blue circle of radius $a$
    (exradius) and it inscribes the yellow circle of radius $b$
    (inradius).  Both circles are centered along the $z$-axis of the
    beam at $x=y=0$.}
  \label{fig1}
\end{figure}
%%%%%%%%%%%%%%%%%%%%%%%%%%%%%%%%%%%%%%%%%%%%

\section{Minimum reconstruction distance}

Let us have a closer look at the minimum reconstruction distance
$z_{0}$, after which a self-reconstructing beam is supposed to restore
its profile. For a single plane wave $\exp(i \vk \cdot \Vr)$, with
wave vector $\vk = k (\hat{x} \sin \theta \cos \phi + \hat{y} \sin \theta \sin
\phi + \hat{z} \cos \theta)$, this parameter can be straightforwardly
estimated in the context of either geometrical and wave
optics~\cite{Aiello:2014aa}.

Actually, let us consider an arbitrary obstruction on the $xy$-plane
with an area $O$.  As sketched in Fig.~\ref{fig1}, for a simply
connected region $O$, we can always find the incircle (the largest
circle inscribed in $O$) and the excircle (the smallest circumscribed
circle), both centered on the beam axis~\cite{Ball:1992aa}. The respective
radii are $b$ (inradius) and $a$ (exradius).  Then, elementary
considerations lead us to~\cite{McGloin:2005aa,Litvin:2009aa}
\begin{equation}
  \label{z10}
  z_0 \propto \frac{a}{\tan \theta} ,
\end{equation}
where the proportionality factor essentially depends on the shape of
the obstruction.

Next, notice that for our single plane wave
\begin{equation}
  \label{z20}
  \frac{1}{\tan \theta}   =    
  \frac{k_z}{( k_x^2 + k_y^2)^{1/2}} =   
  \frac{( k^2 -  k_x^2 - k_y^2 )^{1/2}}{( k_x^2 + k_y^2 )^{1/2}},
\end{equation}
provided that $ k_x^2 + k_y^2 \leq k^2$. This condition is necessary
to maintain $k_z$ real-valued and it limits the applicability of the
equation above to beams whose angular spectrum does not contain
evanescent waves \cite{Mandel:1995aa}.  We can thus regard
$z_0$ as a function of
$\kappa = (k_x^2 + k_y^2 )^{1/2}$ in the $k$-space, namely
\begin{equation}
\label{z30}
  z_0 \sim a \, Z(\kappa) :=   
  a \, \frac{\left( k^2 - \kappa^2 \right)^{1/2}}{\kappa}.
\end{equation}

For an arbitrary beam,  the transverse wave vector $\vka$ has a
density distribution function given by $| \widehat{\Psi} (\vka) |^2$. So, we can
define the minimum reconstruction distance $z_0$  as the expected
value of the function $a \, Z(\kappa)$; namely, 
\begin{equation}
  \label{z40}
  \frac{z_0}{a} = \langle Z(\kappa) \rangle =  
  \frac{\phantom{ix} \displaystyle{
 \iint \frac{\left( k^2 - \kappa^2 \right)^{1/2}}{\kappa}
 |\widehat{\Psi} (\vka)|^2\, \dd \kappa}}
  {\displaystyle{\iint  |\widehat{\Psi} (\vka)|^2 \, \dd \kappa }},
\end{equation}
where both integrals are limited to the disk  $k_x^2 + k_y^2 \leq
k^2$.  We stress that this  formula assigns a definite value of $z_{0}$
to any density  $| \widehat{\Psi} (\vka) |^2$:  self-healing does
occur for any beam.

\section{Gaussian beams}

As the Gaussian beam is the simplest example of a transversally
unbounded diffracting beam, we shall use it as our thread to test the
proposed concepts. We take it to be a Gaussian of waist $w_0$, so it
can be written as
\begin{equation}
\Psi(x,y,z) = \exp ( i k z ) \psi (x,y,z) \, ,
\end{equation}  
with $\psi (x,y,z)$ being the fundamental solution of the paraxial
wave equation: 
\begin{align}
  \label{z120}
  \psi(x,y,z) = \frac{1}{z - i z_R} 
  \exp\left[ 
  i \frac{k}{2} \left(\frac{ x^2 + y^2}{z - i z_R}
  \right)\right] \, ,
\end{align}
and $z_R = k w_0^2/2$ denotes the Rayleigh range. 

%%%%%%%%%%%%%%%%%%%%%%%%%%%%%%%%%%%%%%%%%%%%%%%%%%
\begin{figure*}[t]
  \centerline{\includegraphics[width=\columnwidth]{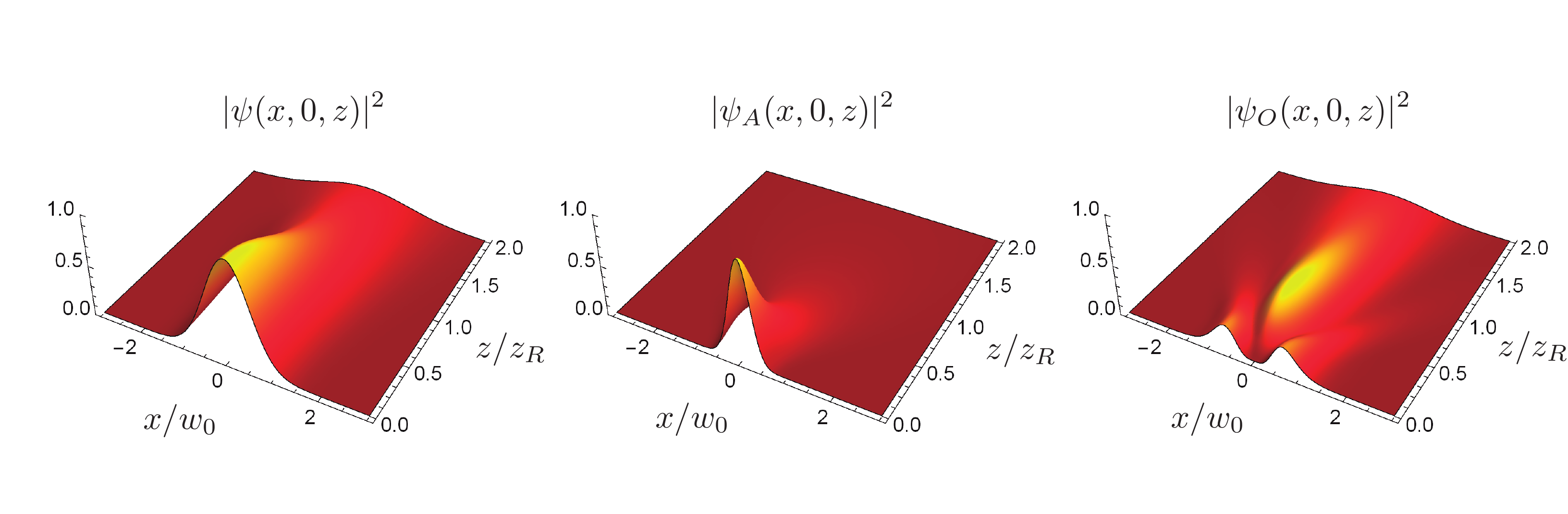}}
  \caption{ \label{fig2} Intensity distributions (evaluated at $y=0$),
    of (from left to right): the incident field $\psi (x,0,z)$, the
    ``virtual'' field transmitted by the aperture complementary to the
    obstruction $\psi_A(x,0,z)$, and the field transmitted behind the
    obstacle $\psi_O(x,0,z)$. The plots correspond to a Gaussian beam
    $w_{0} = 0.26$~mm and a soft-edge Gaussian obstruction with full
    width $a/w_0 = 0.28$.  At   $z/z_{R} = 2$, the intensity profiles
    of  $\psi (x,0,z)$  and $\psi_O(x,0,z)$ appear very similar.}
\end{figure*}
%%%%%%%%%%%%%%%%%%%%%%%%%%%%%%%%%%%%%%%%%%%%%%%%%%

To facilitate the calculations, the obstruction is taken as a
soft-edge Gaussian obstacle of full width $2a$ located along the axis
of the beam at $z=0$.  This is described by the transmission function
\begin{align}
  \label{pz100}
  t_O(x,y) = 1 - \exp\left(-  \frac{| \vrho -\vrho_0|^2}{2 a^2}\right),
\end{align}
where $\vrho_0 = \hat{x} x_0 + \hat{y} y_0$ represents the
displacement of the obstacle with respect to the beam propagation
axis.  The Fourier transformations are straightforward and we finally get the
following expression for the beam transmitted by the virtual aperture
complementary to the obstruction: 
\begin{align}
  \label{z150}
  \psi_A(x,y,z) = \frac{a_R}{z_R}\frac{1}{z - i a_R} 
  \exp\left[  
  i \frac{k}{2} \left(\frac{ x^2 + y^2}{z - i  a_R}\right)\right] \, , 
\end{align}
where, for the sake of clarity, we have chosen $\vrho_0 = 0$ and we
have defined the modified Rayleigh range $a_R$  as
\begin{align}
  \label{z160}
  a_R = \frac{\displaystyle{z_R}}{\displaystyle{1 + \frac{z_R}{k a^2}}} \leq z_R.
\end{align}
The self-healing mechanism of the Gaussian beam is vividly
illustrated in Fig.~\ref{fig2}. A close inspection of this figure
reveals how the self-reconstruction works. From \eqref{z160} it
follows that $a_R \leq z_R$. Therefore, the ``virtual'' field
$\psi_A(x,y,z)$ transmitted by the complementary aperture spreads in
the $xy$-plane, while propagating along the $z$-axis, much more
rapidly than the unperturbed field $\psi(x,y,z)$ and then for
$z/z_{R} \gtrsim 2$ the intensity profile of the obstructed beam
almost coincides with the profile of the unperturbed one.

\begin{figure}[t]
  \centering{\includegraphics[height=4.7cm]{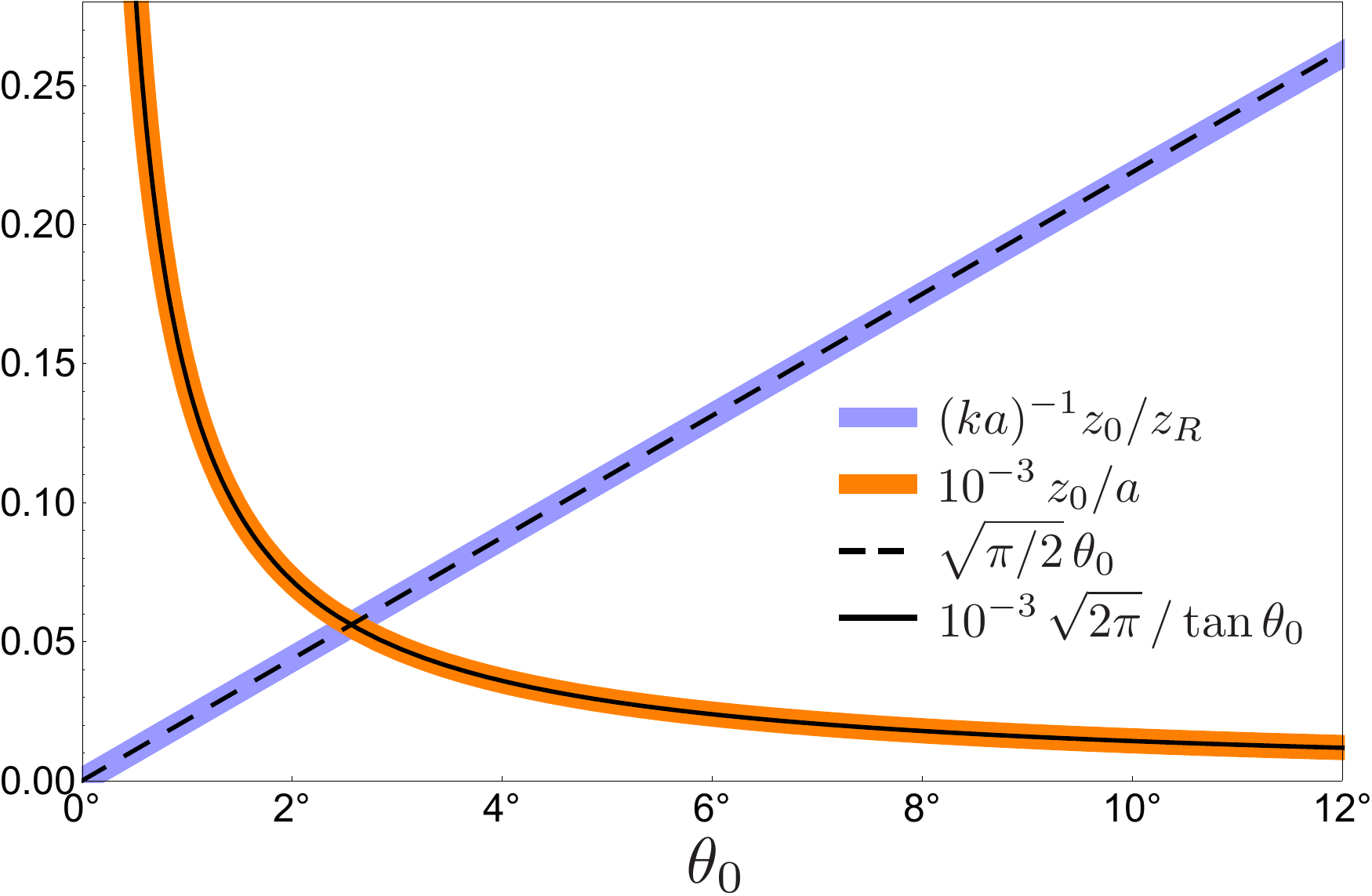}}
  \caption{\label{fig3} Minimum reconstruction distance $z_0/a$ as
    a function of $\theta_{0}$ as well as the paraxial
    approximation. We also plot $z_{0}/z_{R}$, which shows a perfect
    linear behavior. The numerical factor $10^{-3}$ is introduced to
    fit both curves in the same scale. }
\end{figure}

The integrals in \eqref{z40} can be  evaluated analytically; the final
result is
\begin{equation}
  \label{z180}
  \frac{z_0}{a} =  \; \frac{\pi}{2 \theta_0^{\, 2}} \,
  \frac{I_0(1/\theta_0^{\, 2}) + I_1(1/\theta_0^{\, 2})}
  {\sinh ( 1/\theta_0^{\, 2} )} \, ,  
\end{equation}
where $\theta_0 = 2/(k w_0)$ is the angular spread of the Gaussian
beam~\cite{Mandel:1995aa} and $I_\nu(z)$ is the modified Bessel
function of the first kind of order $\nu$. In the paraxial regime,
$\theta_0 \ll 1$ and then
\begin{equation}
  \label{z190}
  \frac{z_0}{a}  \approx  \frac{\sqrt{2\pi}}{\tan \theta_0} \, ,
\end{equation}
which is consistent with the expected geometrical optics result.  A
plot of ${z_0}/{a}$ is given in Fig.~\ref{fig3}, as well as the
paraxial approximation, which works pretty well. Notice that $z_{0}$
is larger for smaller $\theta_{0}$, which might appear
counterintuitive. The reason is that for smaller $\theta_{0}$,
$z_{R}$ becomes larger. To bypass this drawback, we  have also plotted
$z_{0}/z_{R}$, which can be easily obtained from \eqref{z180}. For
$\theta_0 \ll 1$, we get 
\begin{equation}
  \label{z190}
  \frac{z_0}{z_{R}}  \approx ka  \sqrt{\frac{\pi}{2}} \, \theta_{0} \, .
\end{equation}
The goodness of this linear approximation can be appreciated in
Fig.~\ref{fig3}.  

%%%%%%%%%%%%%%%%%%%%%%%%%%%%%%%%%%%%%%%%%%%%%%%%
%%%%%%%%%%%%%%%%%%%%%%%%%%%%%%%%%%%%%%%%%%%%%%%%%%

\section{Quantifying self-healing}

We still have a conundrum pending from the end of Sec.~\ref{sec:eig}:
how is it possible to obtain the simultaneous validity of both
\eqref{p10} and \eqref{p40}?

Indeed, what one really needs is simply to satisfy \eqref{p40} on the
$xy$-plane in the neighborhood of the propagation axis $z$. This
statement may be formalized as follows.  Consider again the
obstruction represented in Fig.~\ref{fig1} that occupies the region
$O$ in the $xy$-plane.  Let $E$ be an arbitrary area in the $xy$-plane
\emph{strictly} contained within $O$. For example, $E$ can be the
region confined by the inner circle of radius $b$, although different
symmetries in the problem may dictate different choices.  Then, as
a necessary condition for self-healing, we require that the amplitude
$\Psi_O(x,y,z)$ of the obstructed beam is proportional to the
amplitude $\Psi(x,y,z)$ of the unperturbed beam \emph{only} within
$E$; viz,
\begin{equation}
  \label{k100}
  \Psi_O(x,y,z ) \Big|_{(x,y) \in E} \approx \lambda_0 \Psi (x,y,z)
  \Big|_{(x,y) \in E} 
  \qquad 
  \forall z \geq z_0 \, .
\end{equation}
From a mathematical point of view, \eqref{k100} makes much more sense
than \eqref{p40}. In fact, the field configuration at $z =0$
completely determines the field distribution at $z>0$. Then, if at a
certain distance $z$, \eqref{p40} were satisfied upon \emph{all} the
$xy$-plane, then it should be also valid at $z=0$. But the latter
statement is clearly false because at $z=0$ one has, by definition,
\sugg{$\Psi_O(x,y,0 ) = t_O(x,y) \Psi (x,y,0) \neq \Psi (x,y,0)$.}
Therefore, the \emph{desideratum} of satisfying both equations
(\ref{p10}) and (\ref{p40}) over all the $xy$-plane cannot be true.

%%%%%%%%%%%%%%%%%%%%%%%%%%%%%% %%%%%%%%%%%%%%%%%
\begin{figure}[t]
 \centering{
 \includegraphics[width= 0.49\columnwidth]{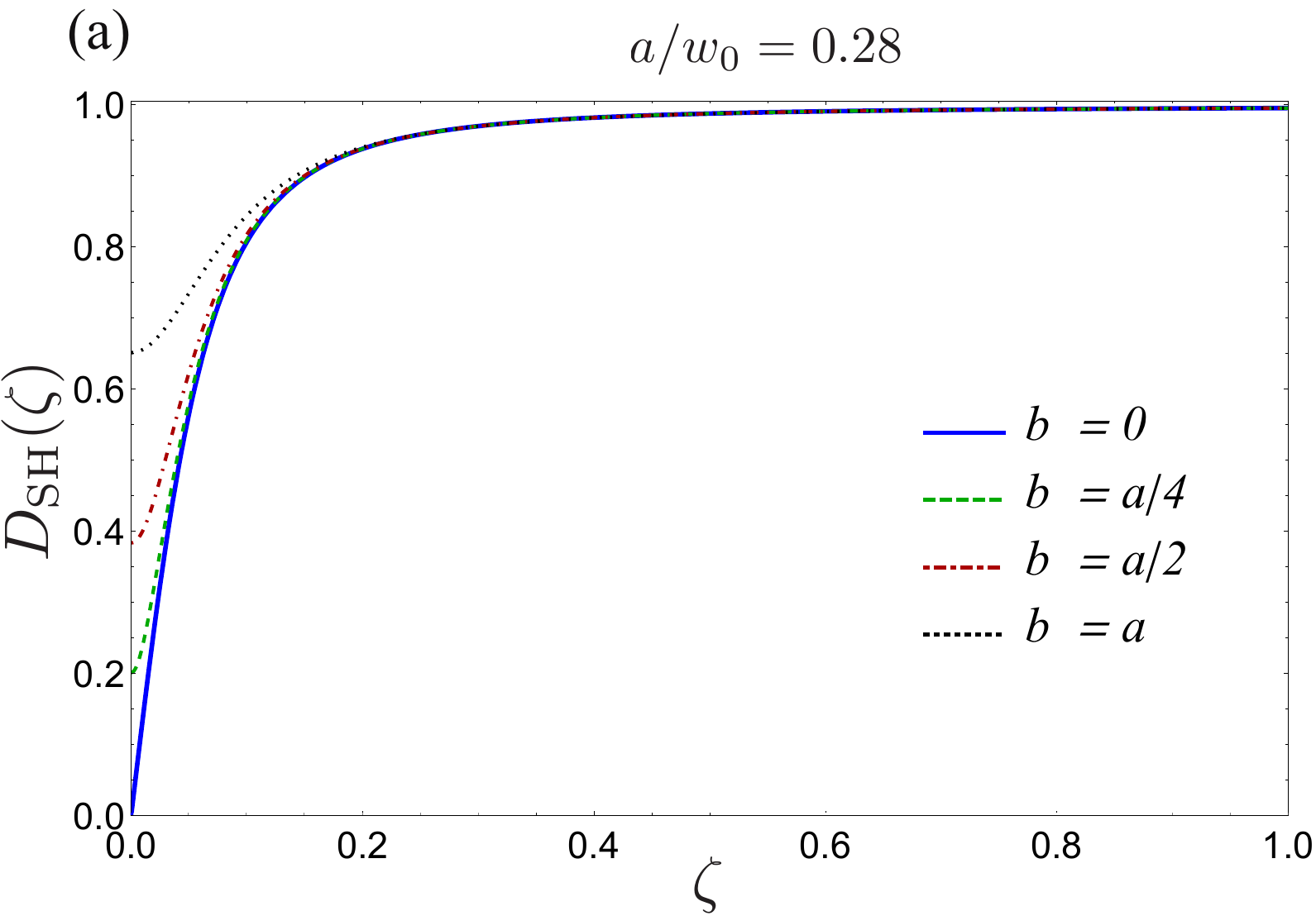}
 \includegraphics[width= 0.49\columnwidth]{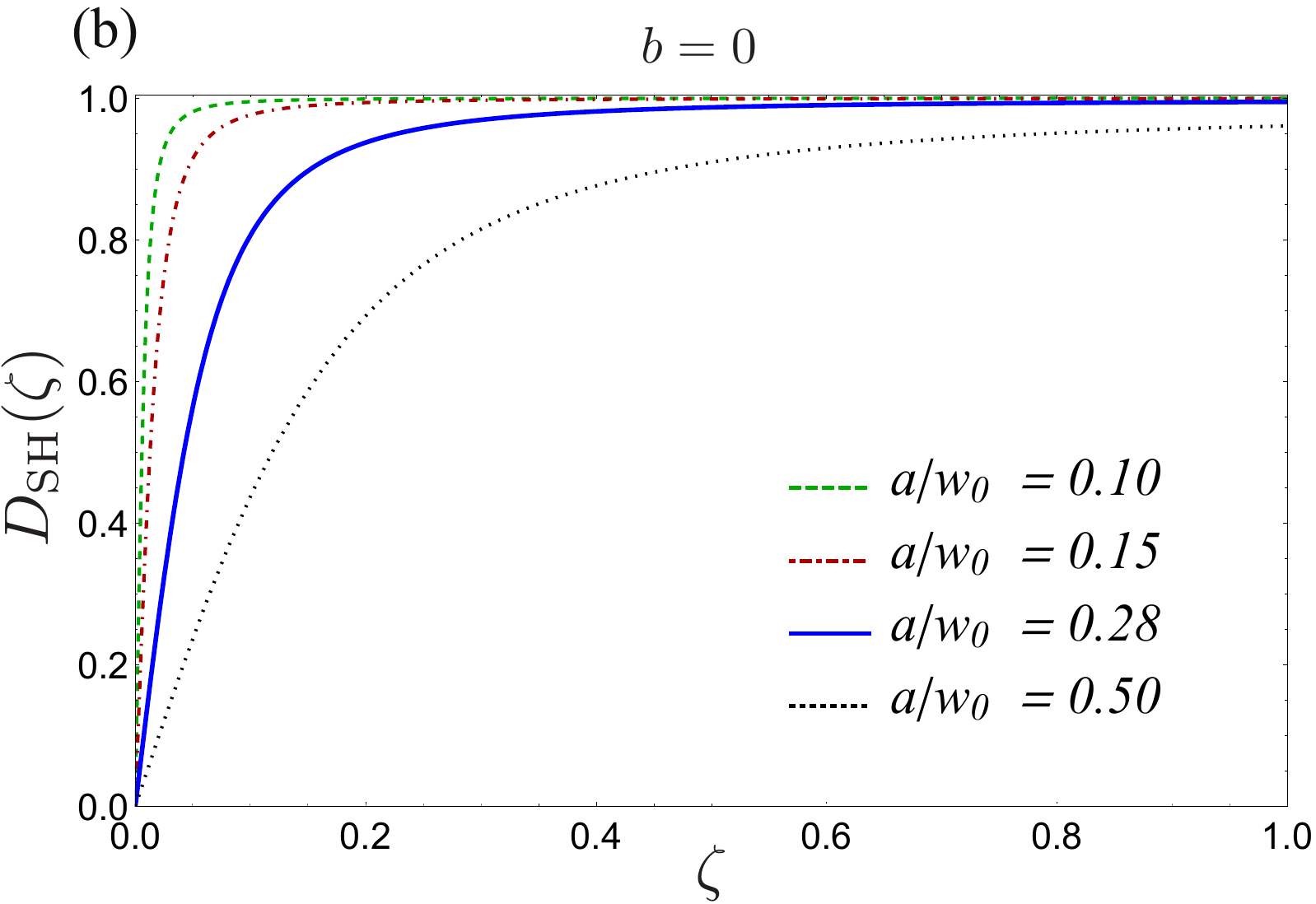}}
  \caption{\label{fig4} (\emph{a}) Plots of the degree of self-healing
    $D_{\mathrm{SH}}  (z)$ for a Gaussian field of waist $w_0$ and different
    radii $b$ of the integration region $E$. The continuous blue line
    represents the limit value for $b \to 0$.  (\emph{b}) The
    limit value of  $D_{\mathrm{SH}}  (z)$  for  $b \to 0$, given in
    \eqref{k170},  for several  values of the width $a$ of the
    soft-edge Gaussian  obstruction.}
\end{figure}
%%%%%%%%%%%%%%%%%%%%%%%%%%%%%%%%%%%%%%%%%%%%%%%%%%

To circumvent this difficulty, we first define a scalar product in
the space of functions $L_2(E)$ as
\begin{equation}
  \label{k120}
  \langle f | g \rangle := \int_{E} f^{\ast} (x,y,z) g(x,y,z)  \, 
  dx  dy \, .
\end{equation}
With this definition, the scalar product $\langle f | g \rangle $ naturally
becomes a function of $z$. Of course, the choice of the integration
domain $E$ is partially discretionary (the only constraint is to be entirely
contained within $O$). However, it is useful to remind here that the
concept of self-healing and minimum reconstruction distance
suffer from the same kind of arbitrariness. In other words, since both
(\ref{p10}) and (\ref{p40}) are impossible to satisfy over the whole
$xy$-plane, one is forced to chose \emph{where} these equations should
be satisfied.  This is because in the total average the field does not
heal.  This follows from Babinet's principle, the perturbation is
somewhere. The beam shape becomes more similar to what it would have
been without obstruction because the effect of the obstruction is
spread out. To some extent this is the core of any self-healing claim
and  defining the healing locally at the position of the obstruction
bypasses the problem.

The scalar product (\ref{k120}) allows us to introduce in a natural
way the corresponding distance $\mathbb{D}(f,g)$ between two functions
in $L_2(E)$ as $\mathbb{D} (f,g) = \norm{f -g}$, where
$\norm{f} = \langle f | f \rangle^{1/2}$.  This distance somehow
quantifies the similarity between the obstructed and the unobstructed
field. In quantum information~\cite{Nielsen:2000aa} there are many
measures of the ``closeness'' of two (normalized) states we want to
compare. Probably, one of the most popular one is the fidelity, a
modified version thereof has been proposed in this context by Chu and
Wen~\cite{Chu:2014aa}. However, the standard fidelity fails to furnish
a quantitative description of self-healing because it is
defined in terms of a scalar product resulting from integration upon
the whole $xy$-plane and this erases any $z$-dependence.

In this paper, we shall use instead the notion of relative distance, which
we define as
\begin{align}
\label{mtpm}
 \mathbb{D}_r(\Psi,\Psi_O)  = 
 \frac{\| \Psi-\Psi_O \|}{\| \Psi+\Psi_O \|} 
  = \frac{\langle \Psi_A | \Psi_A \rangle^{1/2}}
 {\left [\langle \Psi_A | \Psi_A \rangle  +  4 \langle \Psi | \Psi
  \rangle  -  4 \re \langle \Psi | \Psi_A \rangle \right  ]^{1/2}} \, ,
\end{align}
where the scalar products are defined as in \eqref{k120}. 
A direct application of the parallelogram law~\cite{Kolmogorov:1999aa}
[$\norm{f-g}^{2} + \norm{f+g}^{2} = 2(\norm{f}^{2} + \norm{g}^{2})$]
immediately confirms that $0 \le \mathbb{D}_{r}^{2} \le 1$.  If $\Psi_O
\simeq \lambda_0 \Psi$,  with $0 \leq \lambda_0 \leq 1$, then
\begin{align}
\mathbb{D}_r(\Psi,\Psi_O) \simeq  
\frac{1 - \lambda_0}{1+ \lambda_0} \, .
\end{align}

On that account, we find it convenient to introduce a 
$z$-dependent degree of self-healing:
\begin{equation}
  \label{k125}
  D_{\mathrm{SH}} (z) = \sqrt{1 - \mathbb{D}_r^{2} (\Psi, \Psi_{O})} \, , 
\end{equation}
and one can check that $0 \leq D_{\mathrm{SH}} (z) \leq 1$. 
We underline that this concept of distance measure has been
successfully used in assessing a number of key concepts in quantum
optics. \sugg{In general, a distance measure quantifies the extent to
  which two physical states behave in the same way. While these
  distance measures are usually given by certain mathematical
  expressions, they often possess a simple operational meaning, i.e.,
  they are related to the problem of distinguishing the two states.}
The notions of nonclassicality~\cite{Hillery:1987aa},
entanglement~\cite{Vedral:1997aa}, polarization~\cite{Klimov:2005aa}, and
localization~\cite{local}, to cite only a few relevant examples, have
been systematically formulated within this framework. 

For a Gaussian beam, with cylindrical symmetry about the
propagation axis $z$,  we can choose for $E$ a disk of
radius $b \leq a$.  The function $  D_{\mathrm{SH}} (z)$ can be
calculated analytically, although the final expression is complicated and of
little use for our purposes. When $b$ goes to zero, we obtain the
asymptotic form 
\begin{align}
\label{k170}
  D_{\mathrm{SH}} (\zeta) = \zeta 
 \sqrt{\frac{1-\beta^2}{\beta^2 + \zeta^2}}  \, ,
\end{align}
where we have used the dimensionless variables
\begin{equation}
\zeta = \frac{z}{z_R} \, , \qquad
\alpha = \frac{a}{w_{0}} \, , 
\end{equation}
and $\beta = \alpha^2/(1 + \alpha^2)$.  It is interesting to
notice that this function does not depend explicitly on the angular
spread $\theta_0$ of the Gaussian beam.  

%%%%%%%%%%%%%%%%%%%%%%%%%%%%%%%%%%%%%%%%%%%%%%%%%%
\begin{figure*}[t]
  \centerline{\includegraphics[width=\columnwidth]{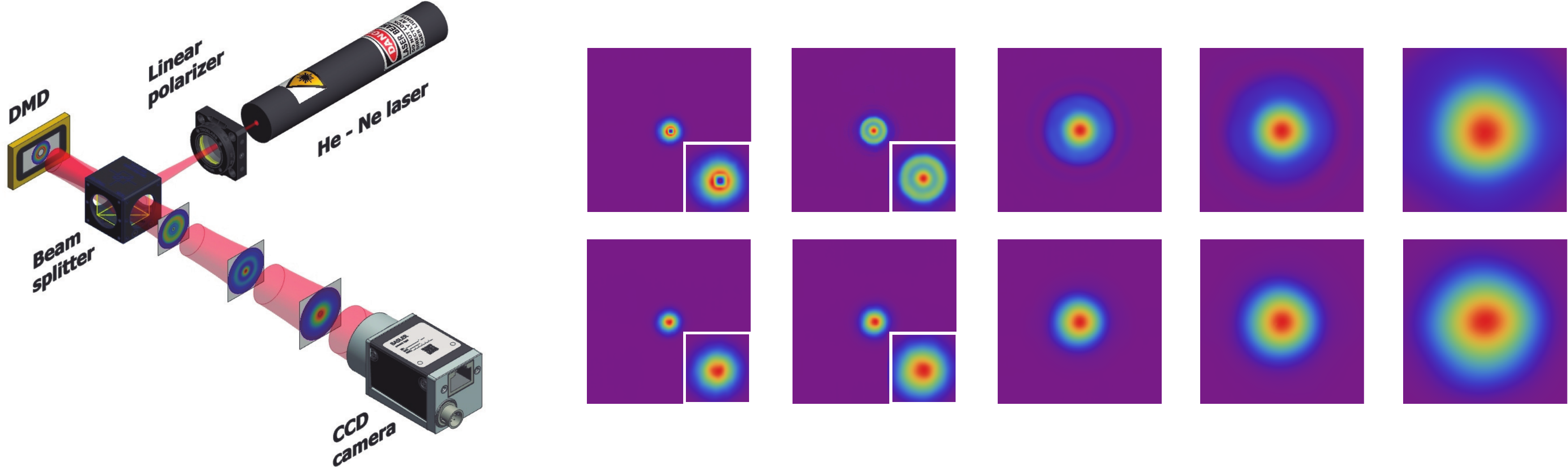}}
  \caption{ \label{fig5} (Left panel) Experimental setup used to check
    the self-healing of a fundamental Gaussian beam created by the
    He-Ne laser. (Right panel) Intensity scans recorded by the CCD
    camera at increasing distances $\zeta = 0, 0.5, 1.5, 4$ and $6.5$
    (from left to the right). The beam has a waist $w_{0} = 0.24$~mm,
    divergence $\theta_{0} = 0.84$~mrad, and Rayleigh range
    $z_{R} = 285$~mm. The upper row corresponds to the obstructed beam
    (with $\alpha = 0.206$), whereas the lower row is for the
    unobstructed beam.  In the first two scans, the images are very
    small, so we have included insets (in white frames) with enlarged
    pictures to better appreciate the patterns.}
\end{figure*}
%%%%%%%%%%%%%%%%%%%%%%%%%%%%%%%%%%%%%%%%%%%%%%%%%%

In Fig.~\ref{fig4}(a) we plot $ D_{\mathrm{SH}} (\zeta)$, for a fixed value
of $\alpha$, and different radii $b$ of the integration region
$E$. When $b$ increases, the dependence of $ D_{\mathrm{SH}} (\zeta)$ with
$\zeta$ becomes weaker.  In Fig.~\ref{fig4}(b) we plot the limit form
of $ D_{\mathrm{SH}} (\zeta)$, given in \eqref{k170}, for different values
of $\alpha$. When $\alpha$ goes to zero, $ D_{\mathrm{SH}} (\zeta)$ tends
to the unity, as expected from physical considerations.

\section{Experiment}

We have checked these predictions in the laboratory. To build up a
Gaussian beam with a central obstruction, a He-Ne laser beam (633~nm,
Thorlabs) was used. The beam impinges on a digital micromirror device
(DMD) chip (Texas Instrument),  with square micromirrors of
7.6~$\mu$m size each.  The obstruction  was generated as an off-state
region on this chip. A sketch of the setup is presented in
Fig.~\ref{fig5}.  All the previous treatment can be directly applied 
to this reflection mode.

%%%%%%%%%%%%%%%%%%%%%%%%%%%%%%%%%%%%%%%%%%%%%%%%%%%
\begin{figure}[t]
  \centerline{\includegraphics[width=0.88\columnwidth]{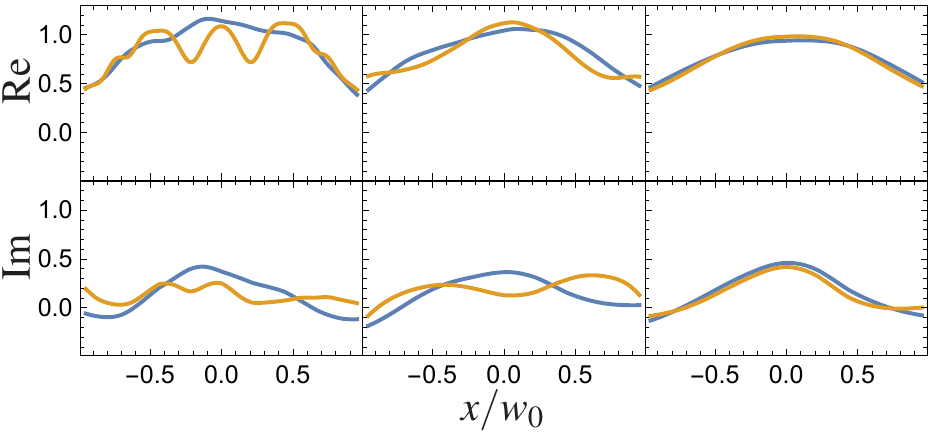}}
  \caption{Real and imaginary parts of the energy-normalized field
    amplitudes at the     positions $\zeta = 0.05, 0.28,$ and 0.56
    (from left to right). The   obstructed field is represented in
    orange, while the unobstructed  is in blue. The obstruction is
    characterized by $\alpha = 0.14$.} 
  \label{fig6}
\end{figure}
%%%%%%%%%%%%%%%%%%%%%%%%%%%%%%%%%%%%%%%%%%%%%%%%%%%
%%%%%%%%%%%%%%%%%%%%%%%%%%%%%%%%%%%%%%%%%%%%%%%
\begin{figure}[b]
  \centerline{\includegraphics[height=4.8cm]{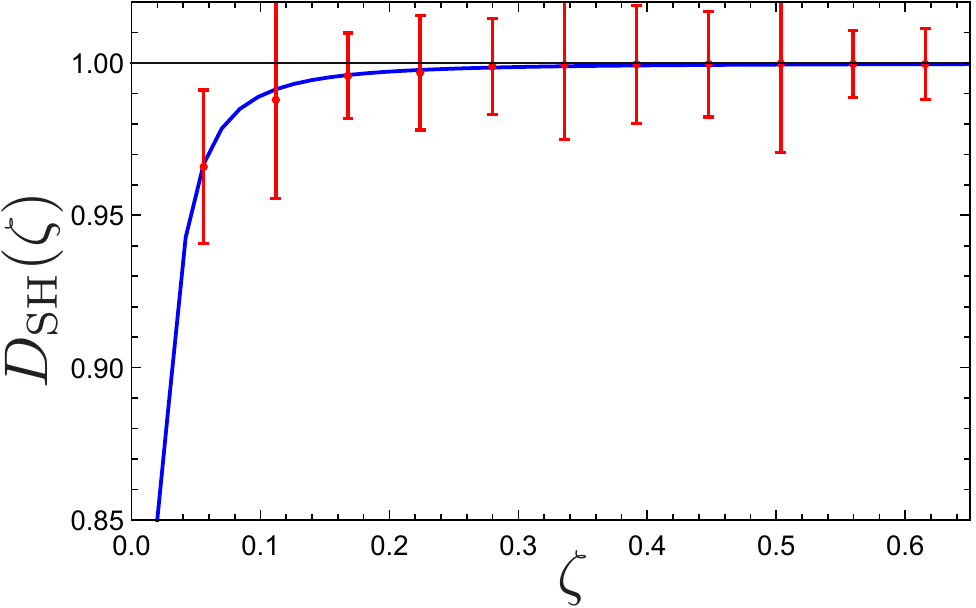}}
  \caption{Experimentally determined degree of self-healing
    $D_{\mathrm{SH}} ( \zeta)$ obtained from the field measurements
    shown in Fig.~6. The integration region $E$ is a dist of radius
    $b=a=0.07~\text{mm}$. The error bars represent standard deviations.}
  \label{fig7}
\end{figure}
%%%%%%%%%%%%%%%%%%%%%%%%%%%%%%%%%%%%%%%%%%%%%%%%%%% 

First, we observed the intensity self-reconstruction of a Gaussian
beam of waist $w_{0} = 0.24$~mm, divergence $\theta_{0} = 0.84$~mrad,
and Rayleigh range $z_{R} = 285$~mm.  The beam was propagated a
distance $z = z_{R}$, where the half-width is $w_{z_{R}} =
0.34$~mm. Then, \sugg{the DMD is inserted at this position where we generate
a centered obstruction of either circular or square shape of
half-widths $a$ of 0.09~mm.} For both shapes of the obstruction the
results are much the same.  Then, the intensity scans are captured in
several positions by a CCD camera (Basler) with 5.5~$\mu$m pixel size.
Some of these intensity profiles (for the case of a square
obstruction) are depicted in Fig.~\ref{fig5} for different propagating
distances from the obstruction.

To experimentally assess the degree of self-healing
$D_{\mathrm{SH}} (\zeta)$ we must be able to measure the whole complex
amplitude for both the obstructed and the unobstructed fields, as it
is apparent from \eqref{mtpm}. To facilitate the measurement, a
calibrated beam expander was used, so the new waist was $w_{0} =
0.6$~mm and the Rayleigh range $z_{R} = 1787$~mm.  Then we place
alternatively the CCD camera and a Shack-Hartmann wavefront sensor
(consisting of a microlens array with 150~$\mu$m lens pitch) to
the same distance from the DMD and measure the intensity and the
wavefront profile of the beam.  To increase the wavefront measurement
resolution, we used another beam expander coupled directly to the
wavefront sensor.  

The field complex amplitude was then reconstructed from these
measurements that were interpolated to the same resolution.  The DMD
was positioned now at a distance of 560~mm from the waist with
half-width $w_{z} = 0.635$~mm. For this measurement, we use the
obscuration with $\alpha = 0.14$, and detection planes at $\zeta$ in
the range 0.05--0.61.  Some of the resulting amplitudes are shown in
Fig.~\ref{fig6}, where the real and imaginary parts are plotted.

Once the complex amplitudes are experimentally determined, we can
compute the degree $D_{\mathrm{SH}} ( \zeta)$. For this purpose, we take
the integration region $E$ as a disk of radius $b= a =0.09$~mm,
which is the size of the obstruction.  Our experimental results are
presented in Fig.~\ref{fig7}. For each distance, the measurements have
been repeated over 100 times, so we can assign error bars. The
agreement with the theory is pretty good.

\section{Concluding remarks}

\sugg{In summary, we have presented a general theory of the so-called
  self-healing process occurring in diverse partially obstructed
  optical beams, whose validity is not limited to diffractionless
  beams as, e.g., Bessel beams. From a careful analysis of the
  physical mechanisms involved, we could ascertain the minimum
  propagation distance from the obstacle after which an optical beam
  recovers its original intensity profile. Our results, obtained
  within the framework of wave optics, confirm and extend the
  traditional ones based on purely geometrical arguments.}

  \sugg{We have quantified self-healing as the closeness between the
  obstructed and unobstructed beams, proposing a suitable measure that
  has been experimentally tested for Gaussian beams, getting a
  beautiful agreement with the proposed theory.}

\section*{Funding}
Technology Agency of the Czech Republic (TE01020229);  
Grant Agency of the Czech Republic (15-03194S); 
IGA Palack{\'y} University (IGA PrF 2016-005); 
MINECO (FIS2015-67963-P).

\section*{Acknowledgments}
We thank Miguel Alonso  for comments that greatly improved the
manuscript.
G. S. A. thanks the BioPhotonics Initiative of the Texas A\&M
University.

\end{document}